%% file: main_clean.tex
\documentclass[times,twocolumn,final]{elsarticle}

\usepackage{xcolor}
\usepackage{geometry}
\newcommand{\verso}[1]{\def\@verso{#1}}
\newcommand{\snm}[1]{#1}
\newcommand{\KWD}{\vspace*{1em}}

\usepackage{framed,multirow}
\usepackage{amsmath,amssymb,amsfonts}
\usepackage{algorithmic}
\usepackage{caption}
\usepackage{hyperref}
\usepackage{booktabs}

\begin{document}

\verso{Gaggion \MakeLowercase{\textit{et al.}}: Multi-view Hybrid Graph Convolutional Network}

\begin{frontmatter}

\title{Multi-view Hybrid Graph Convolutional Network for Volume-to-mesh Reconstruction in Cardiovascular MRI}

\author[1]{Nicolás \snm{Gaggion}}
\author[2]{Benjamin A. \snm{Matheson}}
\author[2]{Yan \snm{Xia}}
\author[2]{Rodrigo \snm{Bonazzola}}
\author[2]{Nishant \snm{Ravikumar}}
\author[2]{Zeike A. \snm{Taylor}}
\author[1]{Diego H. \snm{Milone}}
\author[3,4,5]{Alejandro F. \snm{Frangi}}
\author[1]{Enzo \snm{Ferrante}\corref{cor1}}
\cortext[cor1]{Corresponding author: eferrante@sinc.unl.edu.ar}
  
\address[1]{Institute for Signals, Systems, and Computational Intelligence, sinc(i) CONICET-UNL, Santa Fe, Argentina}
\address[2]{Centre for Computational Imaging and Simulation Technologies in Biomedicine (CISTIB), School of Computing, University of Leeds, Leeds, UK}
\address[3]{Christabel Pankhurst Institute, The University of Manchester, Manchester, UK}
\address[4]{Centre for Computational Imaging and Modelling in Medicine (CIMIM), Department of Computer Science, School of Engineering, and Division of Informatics Imaging and Data Science, School of Health Sciences, The University of Manchester, Manchester, UK}
\address[5]{Medical Imaging Research Centre (MIRC), Department of Cardiovascular Sciences, KU Leuven, Leuven, Belgium}
      
\begin{abstract}
Cardiovascular magnetic resonance imaging is emerging as a crucial tool to examine cardiac morphology and function. Essential to this endeavour are anatomical 3D surface and volumetric meshes derived from CMR images, which facilitate computational anatomy studies, biomarker discovery, and in-silico simulations. \textcolor{black}{Traditional approaches typically follow complex multi-step pipelines, first segmenting images and then reconstructing meshes, making them time-consuming and prone to error propagation.} In response, we introduce HybridVNet, a novel architecture for direct image-to-mesh extraction seamlessly integrating standard convolutional neural networks with graph convolutions, which we prove can efficiently handle surface and volumetric meshes by encoding them as graph structures. To further enhance accuracy, we propose a multi-view HybridVNet architecture which processes both long axis and short axis CMR, showing that it can increase the performance of cardiac MR mesh generation. Our model combines traditional convolutional networks with variational graph generative models, deep supervision and mesh-specific regularisation. Experiments on a comprehensive dataset from the UK Biobank confirm the potential of HybridVNet to significantly advance cardiac imaging and computational cardiology by efficiently generating high-fidelity meshes from CMR images. \textcolor{black}{Multi-view HybridVNet outperforms the state-of-the-art, achieving improvements of up to $\sim$27\% reduction in Mean Contour Distance (from 1.86 mm to 1.35 mm for the LV Myocardium), up to $\sim$18\% improvement in Hausdorff distance (from 4.74 mm to 3.89mm, for the LV Endocardium), and up to $\sim$8\% in Dice Coefficient (from 0.78 to 0.84, for the LV Myocardium), highlighting its superior accuracy.}
\end{abstract}

\begin{keyword}
\KWD Cardiac Imaging \sep Geometric Deep Learning \sep Hybrid Graph Convolutional Neural Network \sep Volume-to-Mesh
\end{keyword}

\end{frontmatter}

\section{Introduction}

Cardiovascular magnetic resonance (CMR) imaging has become an indispensable tool in the diagnosis, treatment planning, and management of cardiovascular diseases. A critical component of advanced cardiac imaging is the extraction of accurate 3D meshes from CMR images. These meshes serve as the foundation for various applications, including computational simulations \cite{fedele2021polygonal}, biomarker discovery \cite{bonazzola2021image}, and analysis of heart deformation and dynamics \cite{meshunetcardiac2022}.

\begin{figure*}
    \centering
    \includegraphics[width=\linewidth]{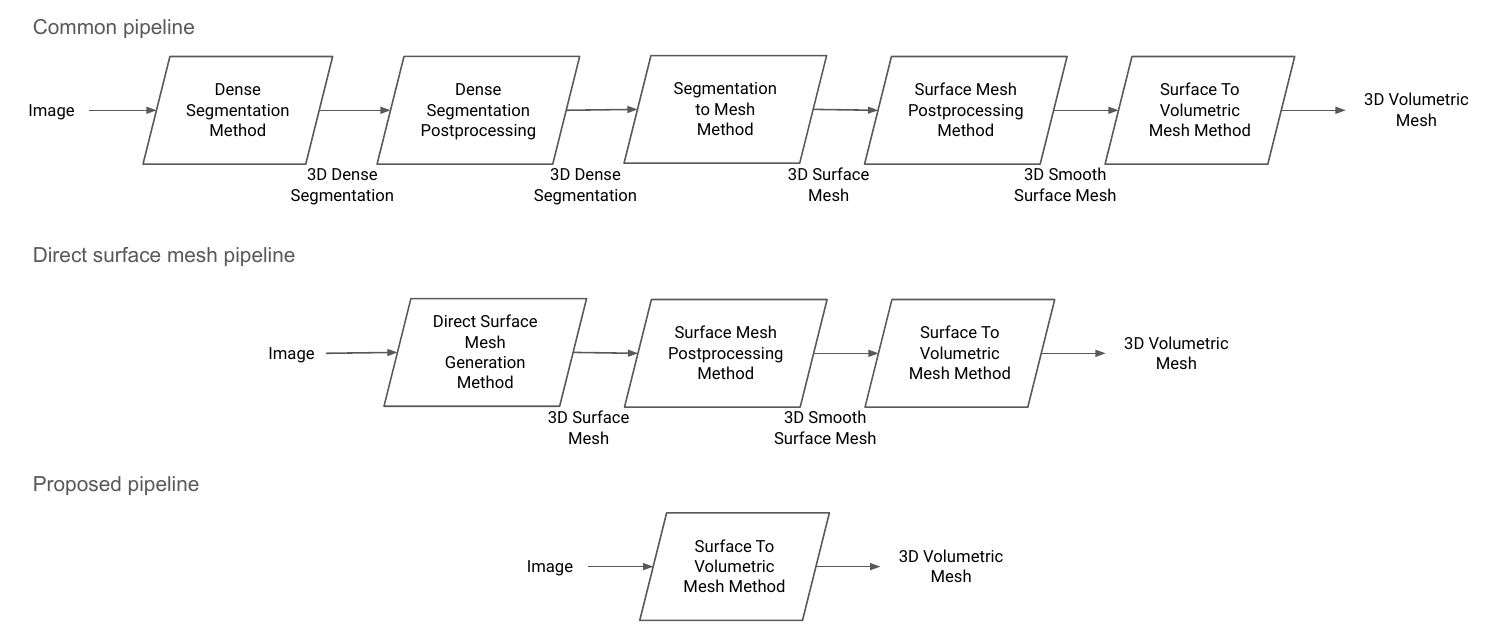}
    \caption{Mesh generation pipelines}
    \label{fig:pipeline}
\end{figure*}

\input{table1}

Despite its importance, cardiac mesh extraction remains a challenging task. Traditional methods, such as active shape models \cite{ordas2007statistical} and multi-atlas segmentation \cite{bai2015multi}, often require extensive computational resources and can be time-consuming. The inherent variability in heart shapes, sizes, and pathologies further complicates the extraction process, necessitating robust and adaptable methods.

Traditional mesh generation pipelines are complex, involving multiple steps and often requiring manual interventions. Figure \ref{fig:pipeline} and Table \ref{tab:mesh_generation} illustrate this complexity, comparing different mesh generation approaches and highlighting the numerous steps, algorithms, and manual interventions typically required in common pipelines.

A particular challenge lies in transitioning from 2D image slices to a cohesive 3D representation, especially when modeling tetrahedral meshes. Current methodologies often require intricate post-processing steps to refine the meshes and make them suitable for simulations \cite{fedele2021polygonal,neic2020automating}. These additional steps can introduce errors and prolong the overall processing time.

\textcolor{black}{Existing approaches to cardiac mesh generation can be broadly categorized into two main strategies. The first strategy follows a multi-stage pipeline that begins with voxel-level segmentation using techniques like U-Net \cite{ronneberger2015u} or V-Net \cite{milletari2016v}, followed by surface mesh extraction and volumetric mesh generation \cite{fedele2021polygonal,neic2020automating, KIM2018161, VAANANEN201919, pak2024robustautomatedcalcificationmeshing}. However, this approach often introduces errors at each stage --segmentation models can produce unrealistic masks with holes or artifacts \cite{larrazabal2020post}, and the subsequent mesh generation steps can compound these errors. Recently, Chen et al. \cite{chen2021shape} proposed MR-Net, which improves this pipeline by using a deep learning-based dense segmentation-to-point cloud registration approach. While MR-Net achieves faster inference times compared to traditional registration methods, it still relies on an initial segmentation step and is therefore limited by the quality of these segmentations. Other methods attempt to improve this pipeline by estimating mesh node displacements \cite{puyol2017multimodal,pak2021} or by deforming a simulation-ready template \cite{kong2021whole}, but their accuracy remains constrained by the quality of the estimated deformations.}

\textcolor{black}{The second strategy aims to bypass these intermediate steps by generating meshes directly from images. Recent work has explored end-to-end neural networks that use convolutional architectures to estimate parameterized shapes \cite{tothova2020probabilistic, XIA2022102498}. While these methods are promising, they typically rely on Principal Component Analysis (PCA) shape models, which are inherently limited by their linear nature and struggle to capture the full complexity of cardiac structures. More recent advances have focused on developing alternative approaches that learn to deform template meshes directly from medical images through various techniques such as differentiable mesh voxelization, graph convolutional networks, and mesh-based motion tracking \cite{JOYCE2022102445, kong2023, meng2023deepmesh}. Although these methods successfully eliminate complex multi-step processing pipelines, they remain constrained by their reliance on template geometries and deformation field estimations, potentially limiting their ability to capture patient-specific anatomical variations.
}

\textcolor{black}{We propose HybridVNet, a novel architecture that advances the direct image-to-mesh approach by combining the strengths of volumetric image processing and geometric deep learning. We work under the hypothesis that direct generation can improve the accuracy of the resulting meshes while being computationally efficient. Motivated by this hypothesis, our method produces high-quality surface and volumetric meshes directly from CMR images through an end-to-end learning approach. Unlike previous methods, HybridVNet uses a hybrid architecture that combines standard 3D convolutions for volumetric image encoding with a spectral graph convolutional decoder for mesh generation. This combination allows us to better capture both global anatomical context and local geometric details, producing meshes that are immediately suitable for computational models without requiring additional processing steps. Notably, while our primary contribution is in direct mesh generation, our experiments also demonstrate that HybridVNet significantly outperforms existing segmentation-to-mesh pipelines, including MR-Net, achieving substantially lower reconstruction errors and better mesh quality.}

\noindent \textbf{Contributions:} Our primary contributions encompass the development of HybridVNet, a multi-view volumetric hybrid graph convolutional model capable of seamlessly integrating multiple CMR views within a jointly learned latent space, directly producing meshes from images. Our model exhibits versatility in creating both cardiac surface and tetrahedral meshes \textcolor{black}{which could potentially be employed} for finite element simulations, \textcolor{black}{both from images or segmentations as input}. We explore classic regularisation techniques for surface meshes and introduce a novel differentiable regularisation term specifically tailored for tetrahedral meshes, markedly enhancing element quality. Notably, while previous works often relied on cropped regions of volumetric images, our model demonstrates exceptional performance in both cropped areas and complete images, showcasing its robustness and adaptability. The performance of HybridVNet is evaluated using the UK Biobank CMR dataset \cite{petersen2015uk}, providing a comprehensive assessment in the context of cardiac imaging. 

\section{Datasets and Reference Meshes}

\subsection{UK Biobank CMR Dataset}

\textcolor{black}{
Data for this study were collected from the UK Biobank (UKB) under access applications 2,964 and 11,350. The study adhered to the guidelines outlined in the Declaration of Helsinki and received ethical approval from the National Research Ethics Service of the National Health Service on 17 June 2011 (Ref 11/NW/0382) and extended on 10 May 2016 (Ref 16/NW/0274). Informed consent was obtained from all participants.}

\textcolor{black}{
The rationale behind the UKB imaging study is explained in \cite{ref_ukbb_cmr}. The UKB resource contains cine-CMR sequences acquired using a balanced steady-state free precession (bSSFP) pulse sequence. The imaging protocol includes a short-axis (SAX) stack covering the full heart with spatial resolution of 1.8 × 1.8 × 10 mm, and three long-axis (LAX) views: two-chamber, three-chamber, and four-chamber views. All sequences were acquired with a temporal resolution of 50 frames per cardiac cycle, as detailed in \cite{petersen2015uk}. For this study, we focused on both end-diastolic (ED) and end-systolic (ES) cardiac phases, representing the points of maximum and minimum ventricular volume in the cardiac cycle, respectively.}

\subsection{Reference Surface Meshes}

\textcolor{black}{
The foundation of our study is a reference cohort of 3D surface meshes introduced by Xia et al. (2022) \cite{XIA2022102498}. These meshes were created through a process of registering a high-resolution atlas of the human heart \cite{rodero2021} to manually delineated 2D contours at ED and ES. The atlas used in this process comprises a mesh that includes six distinct cardiac structures: the left ventricle (LV), right ventricle (RV), left atrium (LA), right atrium (RA), pulmonary artery and the ascending aorta; however, it must be noted that the two latter structures were not present in the manual contours and were inferred from the rest of the structures during registration. The selection criteria for subjects chosen for manual segmentation and the methodology followed are detailed in \cite{petersen2017manualsegm}. Their quality control process included both quantitative and qualitative steps. They first computed the point-to-point distance of the generated shape to the stack of manual contours annotated by medical experts, and if the average error was less than half of the in-plane pixel spacing, then they kept the mesh, otherwise it was discarded. After that, they visually checked all the shapes overlaid on the stack of manual contours to discard any sub-optimal shapes from the dataset. This resulted in the 4525 subjects ultimately included in our study.}

\textcolor{black}{
A key characteristic of this cohort is that each final ground-truth mesh maintains the same number of nodes and set of faces, resulting in an identical adjacency matrix across all meshes. This consistency is a direct result of the atlas registration process and is particularly advantageous for our graph-based approach, as it allows for uniform processing across all samples.}

To ensure fair comparison with baseline methods, we followed the same data splits used in previous works. For image-to-mesh experiments, we used the splits from \cite{XIA2022102498}, with 3925 subjects for training and 600 for testing 
\textcolor{black}{(considering ED and ES phases, giving a total of 1200 meshes)}. Similarly, for segmentation-to-mesh experiments, we adopted the splits from \cite{chen2021shape}, using 957 subjects as the test set \textcolor{black}{(considering ED and ES phases, giving a total of 1914 meshes)}.

\subsection{Volumetric Mesh Generation}

\textcolor{black}{
We derived volumetric mesh ground-truth annotations from the surface meshes through a systematic process that preserves anatomical connectivity. First, we constructed a volumetric atlas mesh using Simpleware software (Version Medical T-2022.03, Synopsys Inc., Mountain View, USA) \cite{Simpleware}. We imported heart structures from the human heart atlas \cite{rodero2021} as individual closed surface meshes of triangular elements. The hollow surface meshes were then populated with tetrahedral elements, ensuring that elements at the interfaces between different cardiac structures share nodes to maintain anatomical connectivity. This resulted in a reference volumetric atlas containing 408,764 elements.}

\textcolor{black}{
To generate subject-specific volumetric meshes, we leveraged the one-to-one correspondence between the surface nodes to register the volumetric atlas to the surface mesh of each subject. This correspondence, inherited from the surface mesh generation process, provides the key landmarks needed for TPS warping \cite{tps_reference}, which was implemented through the Vedo library \cite{vedo_marco_musy_2022_7222019}. TPS warping provides a smooth interpolation between corresponding points while minimizing the bending energy of the transformation, making it particularly suitable for preserving anatomical relationships.}

\begin{figure*}[t!]
    \centering
    \includegraphics[width=\linewidth]{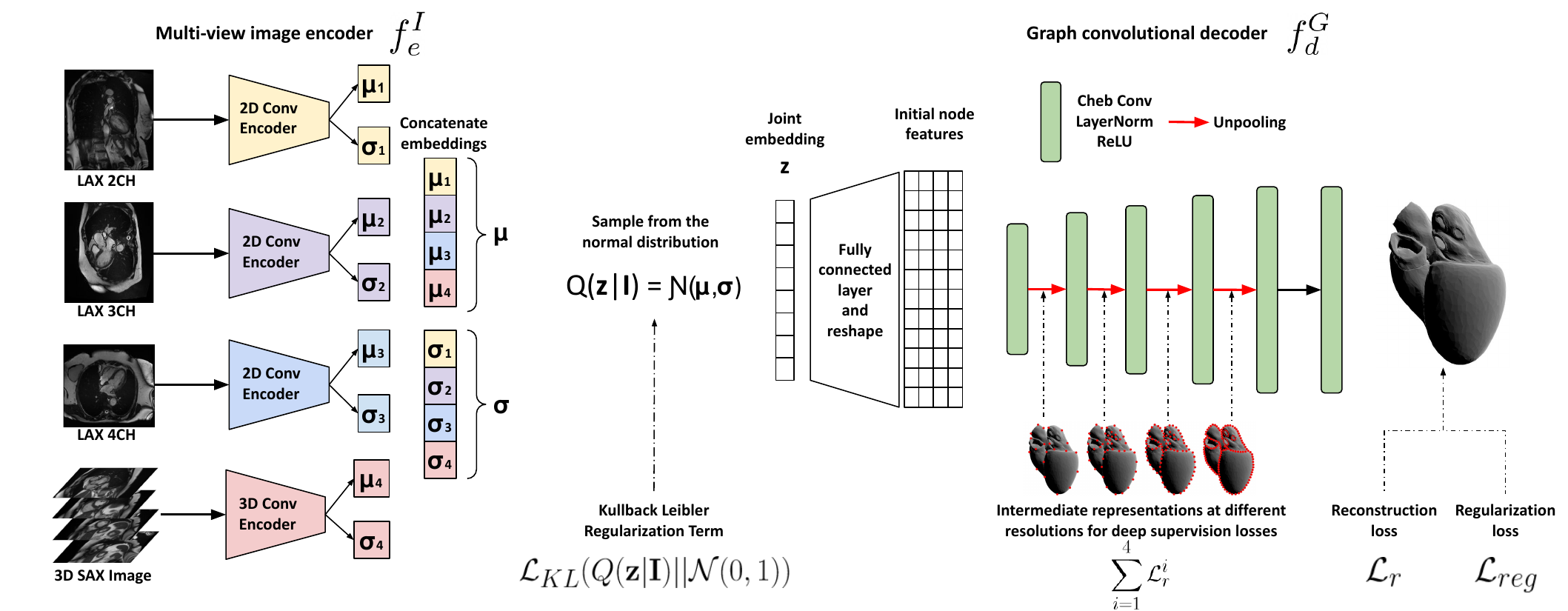}
    \caption{\textbf{Multi-view HybridVNet model architecture:} The proposed model uses a variational encoder-decoder architecture to generate a graph representation of a desired organ from multi-view input images. The encoder consists of independent branches for each input view, concatenated to obtain a joint latent space. The latent code is then passed through a fully connected layer and reshaped to obtain the initial node features for the graph convolutional decoder. This decoder uses the initial node features to generate a final graph representation of the organ.}
    \label{fig:architecture}
\end{figure*}

\section{HybridVNet: Volume-to-Mesh extraction in cardiovascular MR}

In this section we introduce the proposed HybridVNet architecture for volume-to-mesh direction extraction. 
As shown in Figure \ref{fig:architecture}, our HybridVNet model receives multiple CMR views as input: the short-axis view (SAX), which is a 3D cross-sectional view of the heart acquired perpendicular to the long axis, and three different 2D long-axis views (LAX), for two, three and four chambers of the heart (LAX 2CH, LAX 3CH and LAX 4CH, respectively), providing 2D cross-sectional views acquired parallel to the long axis. Given these four images (one volumetric and three 2D), we aim to generate a (surface or tetrahedral) mesh representing the structures of interest.

Consider a dataset $\mathcal{D} = \{(\mathbf{I}, \mathbf{G})_{n}\}_{0 < n \leq N}$, composed of $N$ samples of multi-view CMR images $\mathbf{I} = (\mathbf{I}^{\text{LAX 2CH}}, \mathbf{I}^{\text{LAX 3CH}}, \mathbf{I}^{\text{LAX 4CH}}, \mathbf{I}^{\text{SAX}})$, and their associated cardiac meshes as graphs $\mathbf{G} = \langle V,\mathbf{A},\mathbf{X} \rangle$, where $V$ is the set of $M$ nodes or vertices ($|V|=M$), $\mathbf{A} \in \{0,1\}^{M \times M}$ is the adjacency matrix indicating the connectivity between pairs of nodes ($a_{ij} = 1$ indicates an edge connecting vertices $i$ and $j$, and $a_{ij} = 0$ otherwise), and $\mathbf{X} \in \mathbb{R}^{M \times s}$ is a function (represented as a matrix) assigning a feature vector to every node. It assigns a 3-dimensional spatial coordinate (the mesh vertex position, $s=3$). Since our dataset includes meshes with the same number of nodes and the same connectivity by construction, we can use spectral graph convolutions to decode meshes from a latent space \cite{ranjan2018generating,bonazzola2021image}.

The proposed model consists of a hybrid variational encoder-decoder architecture with multiple inputs. An image convolutional encoder, 
learns a latent representation of the input images, and a spectral graph convolutional decoder
generates a graph representation of the organ. Since our input consists of four images with varying shapes and views, we use a multi-view encoder to handle it. To this end, independent encoder branches are defined for each image view, and a joint latent space is constructed by concatenating their outputs. For all types of LAX images, we use 2D convolutional encoders, $f^{\text{LAX 2CH}}_{e}$, $f^{\text{LAX 3CH}}_{e}$ and $f^{\text{LAX 4CH}}_{e}$, with residual convolutions \cite{he2015deep}. For the 3D SAX image, we use a 3D convolutional encoder, $f^{\text{SAX}}_{e}$, consisting of 3D residual blocks interleaved by max-pooling operations. 

Consequently, our model uses a variational encoder-decoder architecture to generate a graph representation of a desired organ from multi-view input images. The encoder maps the input to a lower-dimensional embedding which represents the parameters of a latent distribution, $\mathbf{z} = f^{I}_{e}(\mathbf{I}^{\text{LAX 2CH}}, \mathbf{I}^{\text{LAX 3CH}}, \mathbf{I}^{\text{LAX 4CH}}, \mathbf{I}^{\text{SAX}})$. \textcolor{black}{This latent distribution is sampled in training by using the reparametrisation trick \cite{kingma2013auto} to ensure a smooth latent space, while at inference time we directly take the mean of the distribution. The sampled latent vector is then passed through a fully connected layer}, and reshaped to obtain initial node features for the graph convolutional decoder, $f^{G}_{d}$. Following the variational autoencoder formulation, the latent code is assumed to be sampled from a multivariate Gaussian, $Q(\mathbf{z} | \mathbf{I}) = \mathcal{N}(\mathbf{\boldsymbol{\mu}}, \text{diag}(\boldsymbol{\mathbf{\sigma}}))$. The distribution is parameterised by the concatenation of outputs from the joint multi-view encoder, $({\boldsymbol{\mu}}, \boldsymbol{\mathbf{\sigma}}) = f^{I}_{e}(\mathbf{I})$. Given a sample of the latent code, $\mathbf{z}$, the graph representation of the organ can be obtained through the decoder $f^{G}_{d}(\mathbf{z})$. 

The model is trained by minimising a loss function defined as

\begin{equation}
    \mathcal{L} = \mathcal{L}_{r}(f_{d}(f_{e}(\mathbf{I})), \mathbf{G}) + \lambda_{KL} \  \mathcal{L}_{KL}\left( Q(\mathbf{z} | \mathbf{I}) || \mathcal{N}(0,1)\right),
\end{equation}

\noindent where the first term is the reconstruction loss based on the mean squared error (MSE) of the vertex positions, the second term imposes a unit Gaussian prior $\mathcal{N}(0,1)$ for the latent posteriors via the KL divergence loss ($\mathcal{L}_{KL}$) and $\lambda_{KL}$ is a weighting factor. 

\subsection{Deeply-supervised spectral graph decoder} 

To generate the graph representation of the target organ, we employed a decoder constructed using spectral graph convolutional neural networks (GCNN). Spectral convolutions are based on the eigendecomposition of the graph Laplacian matrix. In this context, we adopt the spectral convolutions introduced by Defferrard et al. (2016) \cite{defferrard2016convolutional}, which constrain the filters to polynomial filters. This constraint arises from the observation that polynomial filters exhibit strict localisation in the vertex domain, consequently reducing the computational complexity of the convolutional operation. For an in-depth understanding of spectral convolutions, please refer to \cite{defferrard2016convolutional}. 

A spectral convolutional layer operates as standard convolutions applied to images and feature maps. It takes an input feature matrix $\mathbf{X}^\ell$ and produces filtered versions $\mathbf{X}^{\ell+1}$ as output. Our spectral decoder architecture comprises five graph-convolutional layers, each complemented by ReLU nonlinearities with previous Layer Normalisation \cite{ba2016layer}. These layers are strategically interleaved with four fixed graph unpooling layers, allowing the network to learn representations at multiple resolutions. 

We implement the technique outlined by Ranjan et al. (2018) \cite{ranjan2018generating} to obtain these multiple resolutions to construct pairs of pooling and unpooling layers. The process begins by estimating the pooling matrix, achieved through an iterative contraction of vertex pairs while maintaining precise surface error approximations using quadric matrices into the atlas surface mesh. Simultaneously, the unpooling matrix is derived to enable the reversal of the pooling transformation. This process is repeated four times (in the previously pooled version of the atlas), resulting in four sets of pooling and unpooling layers, each reducing and increasing the number of nodes by a factor of two, respectively. Importantly, these pooling and unpooling matrices remain fixed during training, as they are estimated only once for the atlas surface mesh.

To increase our model's performance, we apply the concept of deep supervision \cite{deepsupervision}, which involves supervising the network at various resolution levels. During training, we utilise the estimated pooling operation to obtain down-sampled versions of the ground-truth meshes, enabling us to minimise the reconstruction error at each resolution level. Ultimately, we employ a final graph-convolutional layer, without bias and identity activation function, to predict the final vertex positions.

The incorporation of deep supervision terms leads to the following loss function:

\begin{equation}
    \mathcal{L} = \mathcal{L}_{r} + \lambda_{KL} \mathcal{L}_{KL} + \lambda_{DS}  \sum_{i=1}^{4} \mathcal{L}_{r}^{i},
\end{equation}

\noindent where $\mathcal{L}_{KL}$ is the previously defined KL term, $\lambda_{DS}$ is a weighting factor, and the index $i$ indicates the resolution level of the graph. \textcolor{black}{Best results were obtained with $\lambda_{DS}$ set to 1, giving all different reconstructions the same weight in the final loss.}

\subsection{Mesh regularisation loss functions}

To ensure smooth meshes, state-of-the-art approaches to surface mesh generation often use regularisers such as normal regularisation, edge length regularisation, and Laplacian smoothing ($\mathcal{L}_{lap}$), as introduced in \cite{pixel2mesh}, which we also incorporate. However, these existing metrics were initially designed for triangular surface meshes and, therefore, do not consider the structure of tetrahedral elements in a volumetric mesh \cite{pak2021}. We propose a new regularisation loss function designed to generate tetrahedral volumetric meshes to address this limitation directly. We introduce our new \textit{tetrahedral element regularisation} loss,

\begin{equation}
    \mathcal{L}_{ter} = \frac{1}{N_{t}} \sum^{N_{t}}_{i = 1} \frac{1}{6} 
    \sum^{6}_{j = 1} \left( ||\textbf{e}^{i}_{j}||_{2} - \frac{1}{6}(\sum^{6}_{k = 1} ||\textbf{e}^{i}_{k}||_{2}) \right) ^{2},
    \label{eq:ter}
\end{equation}

\noindent where $N_t$ is the number of tetrahedra, $i$ represents the $i^{th}$ tetrahedron and $\textbf{e}^{i}_{j}$ and $\textbf{e}^{i}_{k}$ represent the edges of that tetrahedron. 
This regularisation term encourages the formation of well-shaped tetrahedral elements by penalizing large variations in edge lengths within each tetrahedron, promoting more uniform and stable volumetric meshes. Such meshes are crucial for accurate finite element simulations in computational cardiology.
The final loss function used to train the model is:

\begin{equation}
    \mathcal{L} = \mathcal{L}_{r} + \lambda_{KL} \mathcal{L}_{KL} + \lambda_{DS}  \sum_{i=1}^{4} \mathcal{L}_{r}^{i} + \lambda_{reg} \mathcal{L}_{reg},
\end{equation}

\noindent where $\mathcal{L}_{reg}$ can be any of the regularisation losses mentioned above: $\mathcal{L}_{lap}$ for the surface case or $\mathcal{L}_{ter}$ for the volumetric case, and $\lambda_{reg}$ is the corresponding weighting factor. \textcolor{black}{We explored combining both regularisation terms for volumetric meshes by applying the Laplacian regularisation solely to surface faces, but this combination showed no improvements when used alongside the tetrahedral regularisation term. The choice between surface or volumetric mesh generation is determined by the adjacency matrix used in the model.}

\section{Experimental Setup}

\input{table2}

\subsection{Image and mesh pre-processing} 

CMR images were pre-processed by normalising intensities to the range [0, 1]. SAX images had dimensions ranging from (100, 100, 6) to (200, 200, 16) and a voxel spacing of [1.82, 1.82, 10] mm, while LAX images had varying dimensions depending on the associated SAX image. To handle different sizes of SAX images between subjects, we evaluated our model in two settings: (1) \textit{Full image input}, where we padded all SAX images to (210, 210, 16), and (2) \textit{Cut input}, where we followed previous work \cite{attar2019quantitative,XIA2022102498} and cropped SAX images to (100, 100, 16), padding slices as needed. In all cases, the LAX images were zero-padded to have a square shape of size (224, 224).

Inspired by classic object detection approaches, we align the vertex positions of the mesh with their relative position inside the SAX image, which is effective when using graph generative models for landmark detection \cite{gaggion2022}. We first remove the origin of the SAX image and divide each direction by the corresponding voxel spacing to obtain the positions in the voxel space. For the full-image pipeline, we add the padding applied to the positions and divide by the image size. For the cropped-image pipeline, we subtract the origin of the bounding box and divide it by the image size. With this, we obtain a \textit{relative positional space} for training the models, with a value of (0.5, 0.5, 0.5) indicating a node in the centre of the SAX image. To evaluate the results, we reversed this operation and recovered the original positions in millimetres. \textcolor{black}{No additional pre-processing was performed on the SAX images.}

\vspace{1em}

\subsection{Data augmentation} 

All models were trained using online data enhancement, including intensity enhancement, random rotations of the SAX images (between -10 and 10 degrees), and arbitrary scaling on the x and y axes. The LAX images were scaled to match the scaling performed in the associated SAX image using each LAX image's respective direction vector. We added a step to randomly choose the cropping centre for the cropped model, ensuring that the entire heart is always inside the region. This helps the model avoid dependence on a perfectly centred crop and is an extra data augmentation step.

\subsection{Model implementation and training details}

All models were implemented in Python using the PyTorch framework \cite{pytorch}. The PyTorch Geometric library \cite{pytorch-geometric} was used for the spectral graph convolutional neural network (GCNN) layers. Hyperparameters were selected through grid search, with the $k$ hop neighbourhood parameter \cite{defferrard2016convolutional} set to 6. We conducted training for 600 epochs using the Adam optimiser with a learning rate of $1\text{E}-4$. The batch size was set to 4, and the weight decay was applied at $1\text{E}-5$. A KL divergence weight factor of $\lambda_{KL}=1\text{E}-5$ was introduced, and a learning rate decline with a factor of 0.99 occurred after each epoch. The 2D and 3D Convolutional Neural Network (CNN) encoders consisted of six residual blocks \cite{resnet-paper}. In 2D encoders, the maxpooling layers were interleaved with these blocks. In 3D encoders, max-pooling was applied on the X and Y axes between each residual block, with Z-axis max-pooling at the third layer. After a grid search hyperparameter selection, the latent representations were obtained using fully connected layers in the encoders, with a dimension of 32 for the 3D encoder and 8 for all 2D encoders. GCNN decoders, in both 2D and 3D models, comprised six layers of Chebyshev convolutions with Layer Normalisation \cite{ba2016layer} and ReLU nonlinearities. Classic surface regularisation losses from the PyTorch3D library \cite{ravi2020pytorch3d} were used. These losses included edge length, normal vector, and Laplacian regularisation terms.

Source code is available at \url{https://github.com/ngaggion/HybridVNet}.

\subsection{Model comparison}

\textcolor{black}{We implemented different single and multi-view variants of the HybridVNet architecture and compared our approach against both direct mesh generation and traditional segmentation-to-mesh pipelines.}

For direct mesh generation, we compared with the Multi-Cue Shape Inference Network (MCSI-Net) \cite{XIA2022102498}, which constitutes the state-of-the-art point distribution model for this dataset. MCSI-Net combines two different networks: a position-inference network that predicts the central coordinates of the mesh and a rotation vector, and a shape-inference network that uses CNN layers to infer the parameters of a point distribution model (PDM) based on PCA. This model uses the same SAX and multiple LAX views as ours, but also incorporates patient metadata information into the PDM learning process, \textcolor{black}{such as demographic data (age, weight, height, and body mass) and cardiovascular-related parameters, including lifestyle factors, blood pressure, and laboratory-derived biological markers (comprising a total of 29 variables in addition to the imaging data)}. On the contrary, our model does not require patient metadata. By comparing our HybridVNet with MCSI-Net, we aim to demonstrate the effectiveness of our approach in generating high-quality cardiac meshes without relying on patient metadata.

\textcolor{black}{A fundamental question in cardiac mesh generation is whether direct mesh estimation from images is more efficient than the traditional pipeline of image segmentation followed by mesh reconstruction. To address this, we also compared HybridVNet against baseline approaches that follow the traditional pipeline: first performing automated image-based segmentation, then reconstructing surface/volumetric meshes from the segmentations via sparse point clouds. For this comparison, we evaluated several state-of-the-art methods representing different approaches to mesh reconstruction from segmentation masks, including PointNet++ \cite{qi2017pointnet++}, PU-Net \cite{yu2018pu}, Pixel2mesh \cite{pixel2mesh}, Coherent Point Drift (CPD) \cite{myronenko2010pointcpd}, Gaussian mixture models (GMMREG) \cite{jian2010robustgmmreg}, and  MR-Net \cite{chen2021shape}, the current state-of-the-art in segmentation-to-mesh reconstruction for this dataset. These methods were applied to point clouds generated from automated segmentations.}

\section{Results and Discussion}

We conducted a comprehensive series of experiments to evaluate the performance of the proposed HybridVNet model alongside the baseline models and their various configurations. These experiments covered surface and tetrahedral volumetric mesh scenarios, including a sensitivity analysis of the proposed regularisation losses. All evaluations were carried out on the same test dataset comprising 600 subjects, as presented in \cite{XIA2022102498}, for the ground-truth meshes associated with this dataset. 

\input{table3}

\subsection{Surface mesh extraction}

\begin{figure*}[]
    \centering
    \includegraphics[width=0.95\linewidth]{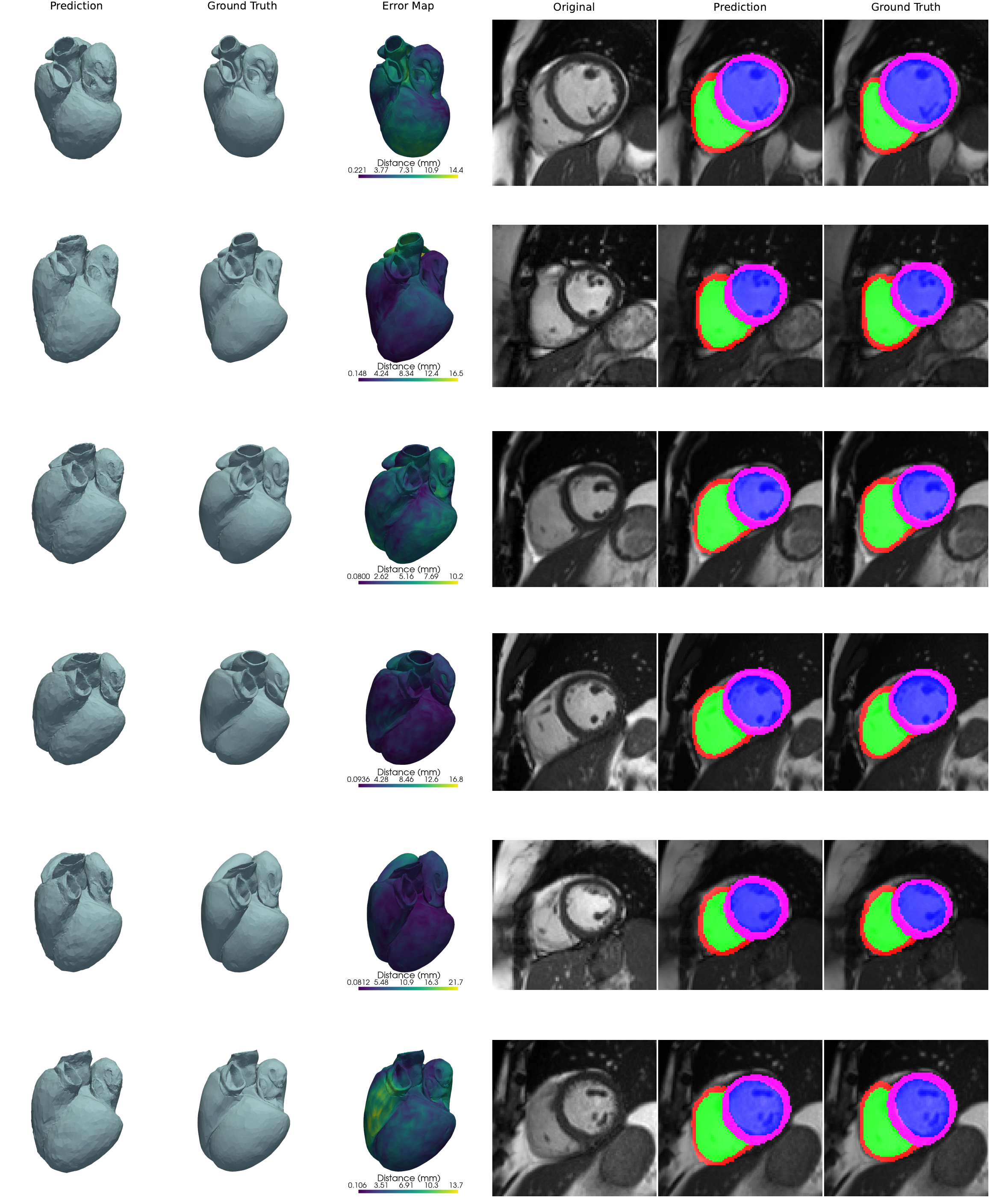}
    \caption{Qualitative performance evaluation of MV-HybridVNet \textcolor{black}{(with $\lambda_{lap}=0.01$)} on cardiac MRI segmentation across six test subjects, showing the predictions against ground truth. For each subject, we present mesh-based visualizations (left three columns) showing the predicted surface, ground-truth surface, and their error map comparison, alongside 2D visualizations (right three columns) \textcolor{black}{at mid-ventricular level displaying the original middle slice MRI}, predicted segmentation overlay, and ground-truth segmentation overlay. The top four rows demonstrate the performance on healthy subjects, while the bottom two rows showcase segmentation results from subjects with myocardial infarction.}
    \label{fig:qualitative}
\end{figure*}

To evaluate the quality of cardiac meshes, we used mesh metrics (Table \ref{surf_mesh_metrics}) and mask-based metrics (Table \ref{surf_seg_metrics}). \textcolor{black}{As shown in Figure \ref{fig:qualitative}, our model generates high-quality surface meshes and accurate segmentations across different test subjects.} First, to enable a direct comparison with MCSI-Net, which was evaluated directly on the segmentation masks generated by the model in the SAX image space, we derived dense segmentation masks from the surface meshes. Then, we evaluated classic segmentation metrics such as Dice coefficient \textcolor{black}{(DC)}, Hausdorff distance \textcolor{black}{(HD)}, and the average distance between the reference and predicted contours in each slice \textcolor{black}{(MCD)}.

In our initial comparison, we evaluated our HybridVNet against the SAX-only MCSI-Net with full images and cropped versions centred on the structure of interest (Table \ref{surf_seg_metrics}). Remarkably, HybridVNet outperforms SAX MCSI-Net for all metrics and structures. Next, we compare our MV-HybridVNet with the standard MCSI-Net, which also incorporates multiple views and is the current state of the art for this data set. The results demonstrate the superiority of our MV-HybridVNet, as it outperforms the standard MCSI-Net across all segmentation metrics for both the left and right ventricle segmentation tasks.

Our \textit{full image} variant of the model achieves better results compared to the baselines, all while eliminating the need for an additional step to detect the region of interest during the segmentation process. Furthermore, the MV-HybridVNet model on \textit{cropped images} beats the results with significant differences relative to the full image. 

To account for structures that may not be visible in SAX images and to provide more insight into how the incorporation of long axis views in our model helps the model learn more details about the complete heart structure, we conducted a thorough evaluation of our proposed models directly on various subparts of the output mesh. Standard mesh evaluation metrics, including vertex mean squared error (MSE) and mean average error (MAE) \textcolor{black}{between reference and predicted surface meshes}, were calculated in millimetres. Table \ref{surf_mesh_metrics} summarises the results in our models, comparing HybridVNet with its multi-view version for \textit{cropped images} and \textit{full images} versions independently. Evaluation was performed at the nodes of the left ventricle (LV), right ventricle (RV), left atrium (LA), right atrium (RA) and aorta.

Comparing the performance of the HybridVNet with and without the inclusion of LAX images, we observed a significant improvement in accuracy for all parts of the mesh. This improvement is particularly pronounced for the left and right atria (LA and RA) and the aorta, which are not fully visible in SAX images. The base HybridVNet model demonstrates the ability to approximate the positions of these structures, with further refinement achieved through the integration of LAX images. \textcolor{black}{Interestingly, note that our method does not require any kind of pre-alignment between LAX and SAX images. Even though we could potentially improve performance even further by doing this pre-alignment, this would imply an additional step, making model use more complicated and prone to errors that may be introduced during the registration process.}

\begin{figure}
    \centering
    \includegraphics[width=\linewidth]{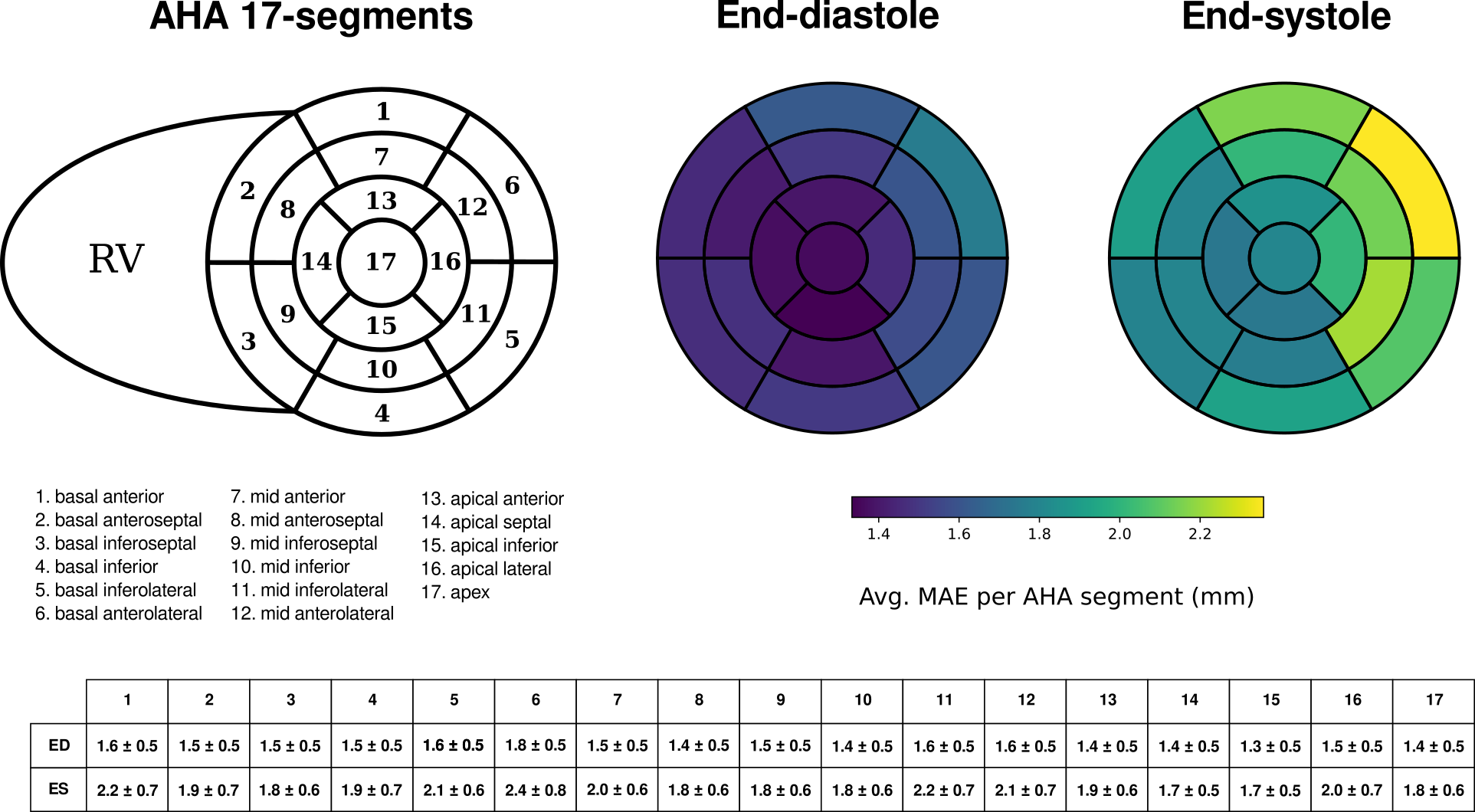}
    \caption{\textcolor{black}{Average segmentation error (measured as MAE) across the 17 AHA left-ventricular segments on the test set of 600 subjects, evaluated at end-diastole (ED) and end-systole (ES). A consistently higher error is observed for ES segmentation, with slightly increased errors around the antero-lateral region in both phases. The table shows the values per segment as average $\pm$ standard deviation.}}
    \label{fig:mae_by_aha_segment}
\end{figure}

\textcolor{black}{Finally, we evaluated the reconstruction performance across different regions of the LV and at the two cardiac phases, ED and ES, using the 17-segment model defined by the American Heart Association (AHA). As shown in Fig.~\ref{fig:mae_by_aha_segment}, reconstruction errors were consistently higher at ES compared to ED, while regional differences were less pronounced overall, with slightly elevated errors observed around the anterolateral region.}

\vspace{1em}
\subsubsection*{Surface mesh regularisation effect} 

In the context of the surface mesh experiment, we performed a comprehensive evaluation of various surface regularisation loss functions to enhance the performance of our HybridVNet model. Specifically, we investigated the efficacy of three distinct regularisation approaches: normal regularisation, edge-length regularisation, and Laplacian smoothing. For more information on these regularisers, see \cite{pixel2mesh}.

\begin{figure}[ht!]
    \centering
    \includegraphics[width=\linewidth]{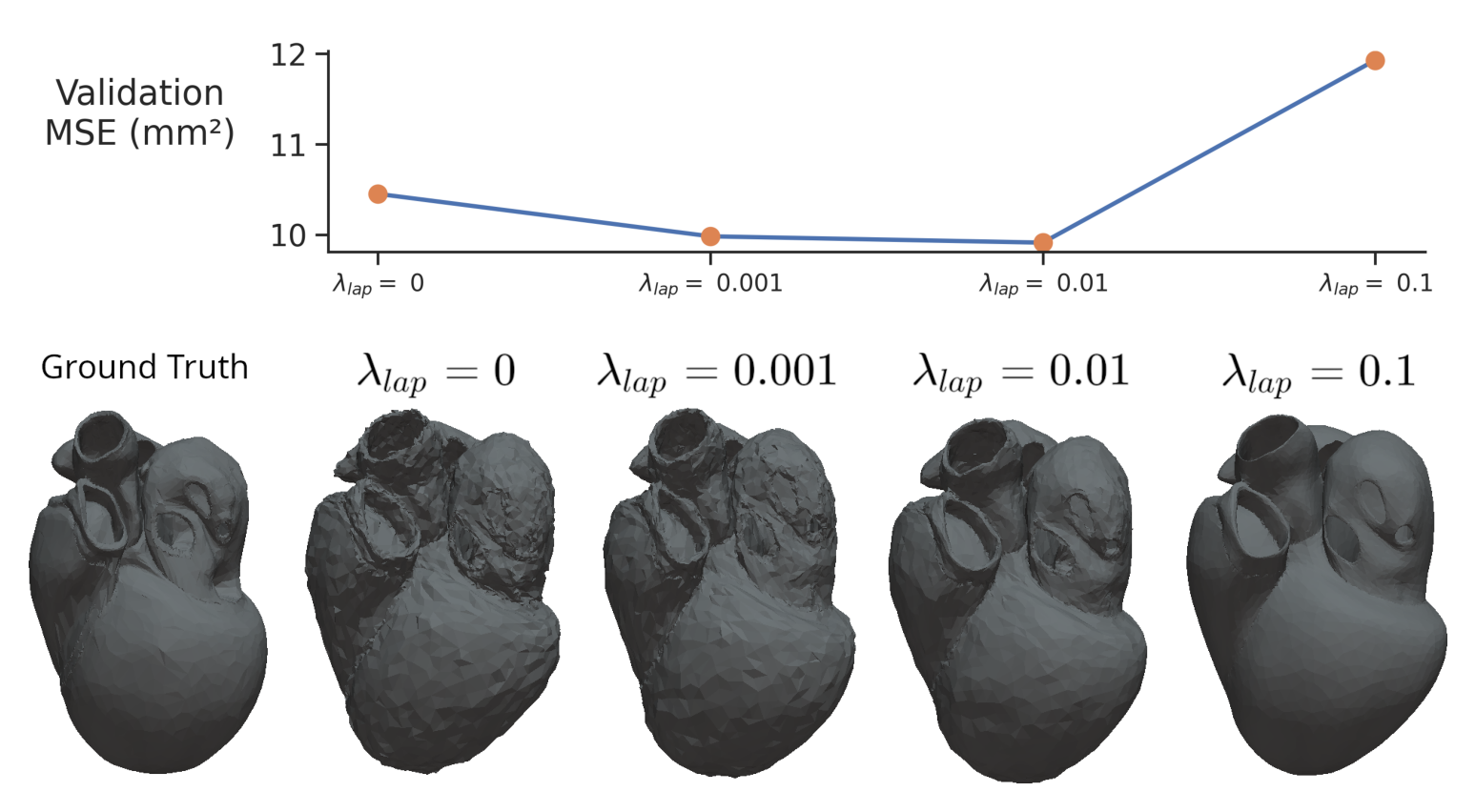}
    \caption{Qualitative analysis of the impact of Laplacian regularisation term on surface mesh smoothness. It demonstrates the influence of adjusting the regularisation parameter on mesh quality. The best quantitative results regarding MSE for the validation split were achieved when $\lambda_{lap} = 0.01$.}
    \label{fig:laplacian_qualititive}
\end{figure}

\begin{figure}[ht!]
    \centering
    \includegraphics[width=\linewidth]{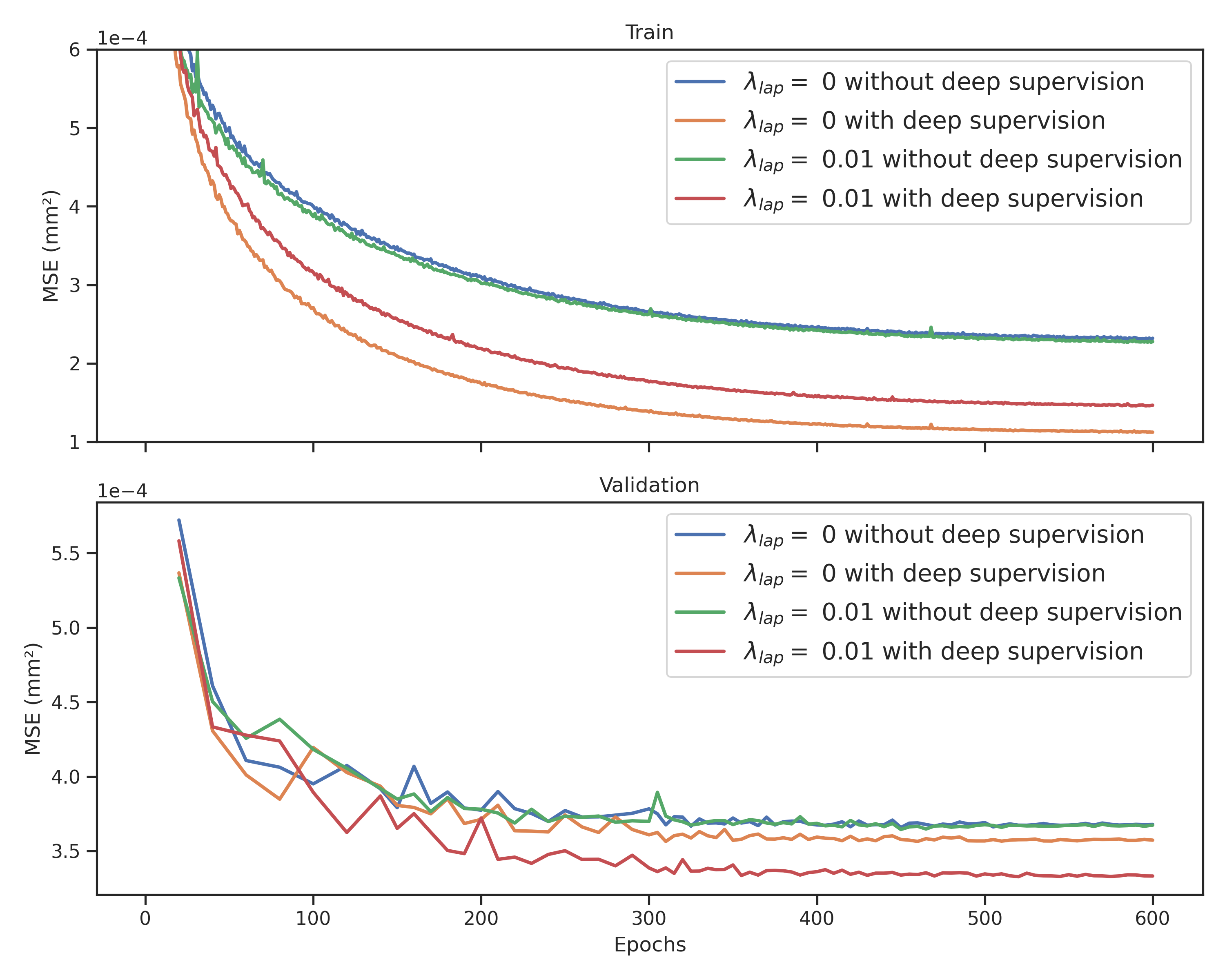}
    \caption{MSE values throughout training and validation for different configurations of hyperparameters, measured in the \textit{relative positional space}. The red curve highlights the significant impact of combining deep supervision and Laplacian regularisation losses on model performance. Smaller intervals were used for loss recording as training progressed.}
    \label{fig:Loss functions}
\end{figure}

Notably, while commonly employed in mesh regularisation tasks, normal regularisation, and edge length regularisation did not yield significant improvements in our model's performance. This observation aligns with the intuitive understanding that these metrics are better suited for meshes with varying node counts and highly irregular target shapes. This is not the case in our dataset. In contrast, the incorporation of Laplacian smoothing produced notably smoother surface meshes. This can be visually appreciated in Figure \ref{fig:laplacian_qualititive}, which presents a qualitative analysis of the meshes obtained as the regularisation parameter for the Laplacian regularisation loss was increased. Figures clearly illustrate the enhanced smoothness and quality of the meshes as the regularisation strength is adjusted.

To assess the impact of different loss terms during the training process, we refer to Figure \ref{fig:Loss functions}. This figure provides a comparison of the MSE values throughout both the training and validation phases. Notably, due to the resource-intensive nature of the validation process, we adjusted the intervals when recording loss values, with smaller intervals as more training time elapsed. 

Significantly,  the red curve in Figure \ref{fig:Loss functions} illustrates that the best performance is achieved when combining both deep supervision and Laplacian regularisation losses. This combination eases the training process and leads to improved model performance. The optimal regularisation strength for Laplacian smoothing, resulting in the best MSE for the entire image and cropped models, was determined to be $\lambda_{lap} = 0.01$. This finding was consistent with both qualitative and quantitative evaluations, as over-smoothed meshes appeared when using high values of the regularisation term.

\subsection{Clinical Validation}

\begin{figure*}[h!]
    \centering
    \includegraphics[width=0.95\linewidth]{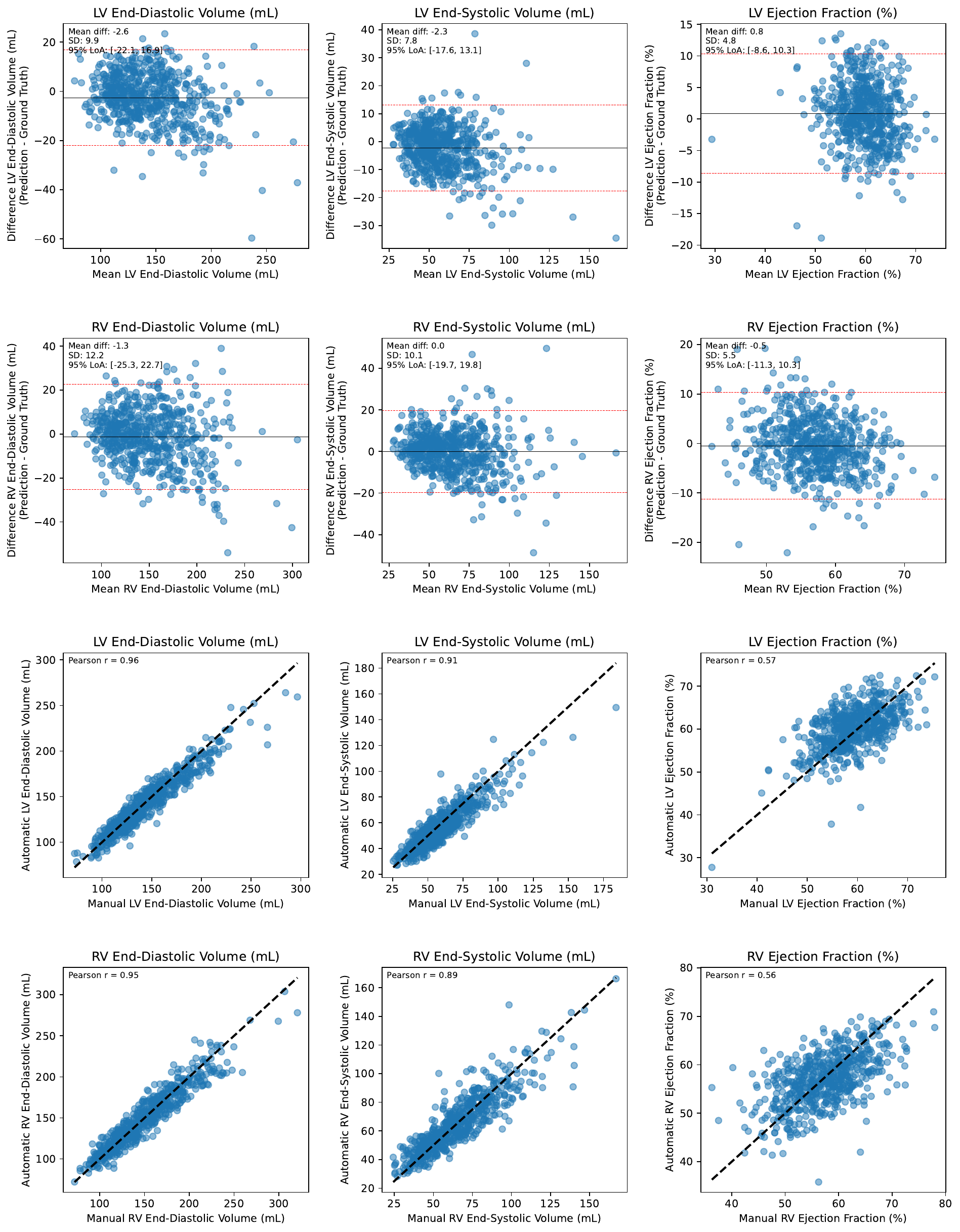}
    \caption{\textcolor{black}{Bland-Altman and Correlation analyses} comparing automated measurements derived from meshes generated with HybridVNet and manual measurements of cardiac parameters across 600 test subjects. Top two rows: Bland-Altman plots showing the agreement between automated and manual measurements, with mean difference (solid line) and 95\% limits of agreement (red dashed lines). Bottom two rows: Correlation plots between automated and manual measurements.}
    \label{fig:clinical_validation}
\end{figure*}

\textcolor{black}{
We performed a clinical validation of our automated cardiac analysis method against manual expert measurements across 600 test subjects, following the procedure in \cite{XIA2022102498}. To derive the clinical cardiac functional indices, we first extract contours corresponding to the intersection between the 3D triangular meshes obtained by our best MV-HybridVNet model and the CMR image slices. Following standard clinical practice, ventricular volumes are calculated using the method of disks, where the total cardiac volume is approximated by summing the areas within 2D segmentation contours and multiplying by the inter-slice spacing.}

\textcolor{black}{Figure \ref{fig:clinical_validation} presents the correlation and Bland-Altman analyses for key ventricular parameters. Our method demonstrated good agreement with manual measurements for ventricular volumes, achieving high correlation coefficients for both left and right ventricles (LVEDV: r=0.957, LVESV: r=0.905, RVEDV: r=0.949, RVESV: r=0.886). The Bland-Altman analysis revealed minimal systematic bias in volume measurements (LVEDV: -2.62±9.94 mL, LVESV: -2.28±7.83 mL, RVEDV: -1.31±12.24 mL, RVESV: 0.02±10.07 mL). For ejection fractions, while the correlations were moderate (LVEF: r=0.572, RVEF: r=0.565), the mean differences were minimal (LVEF: 0.84±4.82\%, RVEF: -0.47±5.50\%), indicating good clinical agreement.}

\subsection{Comparison with Segmentation-to-Mesh Pipeline}

\input{table4}

\textcolor{black}{As shown in Table \ref{tab:comparison_seg_to_mesh}, HybridVNet significantly outperforms the existing segmentation-to-mesh pipelines across all metrics. We considered baseline methods reported in the literature which achieve Chamfer distances (CD) ranging from 12.10 to 20.90 mm and Hausdorff distances (HD) from 13.05 to 18.57 mm. MR-Net, the current state-of-the-art method, achieved a CD of 4.39 ± 1.48 mm and HD of 6.89 ± 1.88 mm. In contrast, HybridVNet improves upon these results with a surface mesh CD of 4.13 ± 1.16 mm and HD of 5.17 ± 1.02 mm, and an even more impressive volumetric mesh CD of 2.80 ± 2.40 mm and HD of 6.09 ± 2.09 mm.}

\textcolor{black}{These results demonstrate that direct mesh estimation from dense segmentation masks using HybridVNet is not only more efficient in terms of computational pipeline but also significantly more accurate than the traditional segmentation-to-mesh approach. This superior performance can be attributed to the ability of HybridVNet to learn end-to-end shape features directly from the image data, avoiding error accumulation that occurs in multi-stage pipelines (in this case, the step of obtaining point-clouds from dense segmentation masks). By eliminating intermediate processing steps, HybridVNet reduces computational overhead and minimizes potential sources of error, resulting in more precise mesh reconstructions.}

\subsection{Generation and quality improvement of tetrahedral meshes}

Our last experiment focused on the creation of tetrahedral meshes, \textcolor{black}{which could potentially be used for simulations, specifically examining the trade-off between mesh quality and anatomical accuracy}. We evaluated various weighting factors ($\lambda_{ter}$) for the regularisation term defined in (\ref{eq:ter}), to understand its influence on both mesh quality and segmentation accuracy. Table \ref{table:volumetric} presents results for different $\lambda_{ter}$ values. 

\input{table5}

To assess mesh quality comprehensively, we employed several standard metrics. The primary metric used was the scaled Jacobian, widely adopted in the field. The Jacobian of a tetrahedron is a matrix that describes how the tetrahedron's shape changes under deformation. The scaled Jacobian provides a quantitative measure of regularity and symmetry, falling within the range [-1, 1] and not affected by scale or units. A high-scaled Jacobian value implies high regularity, low distortion, and therefore high quality \cite{johnen2018efficient}. \textcolor{black}{As a complement, we also evaluated additional tetrahedral quality metrics, which capture different aspects of element quality: aspect ratio indicates elongation, mean ratio assesses deviation from equilateral shape, skewness measures angular deformation, and shape quality evaluates overall element regularity. As shown in Table \ref{table:volumetric}, all quality metrics improve as the regularisation strength increases, with $\lambda_{ter}=1\text{E-}2$ achieving the best absolute scores.}

\textcolor{black}{Our exploration reveals that $\lambda_{ter}$ directly mediates the balance between segmentation accuracy and mesh quality. Lower values ($\lambda_{ter} = 1\text{E-}4$) optimize segmentation performance, closely matching the non-regularized model. In contrast, $\lambda_{ter} = 1\text{E-}3$ provides the best compromise between maintaining anatomical accuracy while achieving acceptable mesh quality for most applications. Higher values ($\lambda_{ter} = 1\text{E-}2$) further improve element quality but at the cost of reduced accuracy in vertex positions.}

\begin{figure}[ht!]
    \centering
    \includegraphics[width=\linewidth]{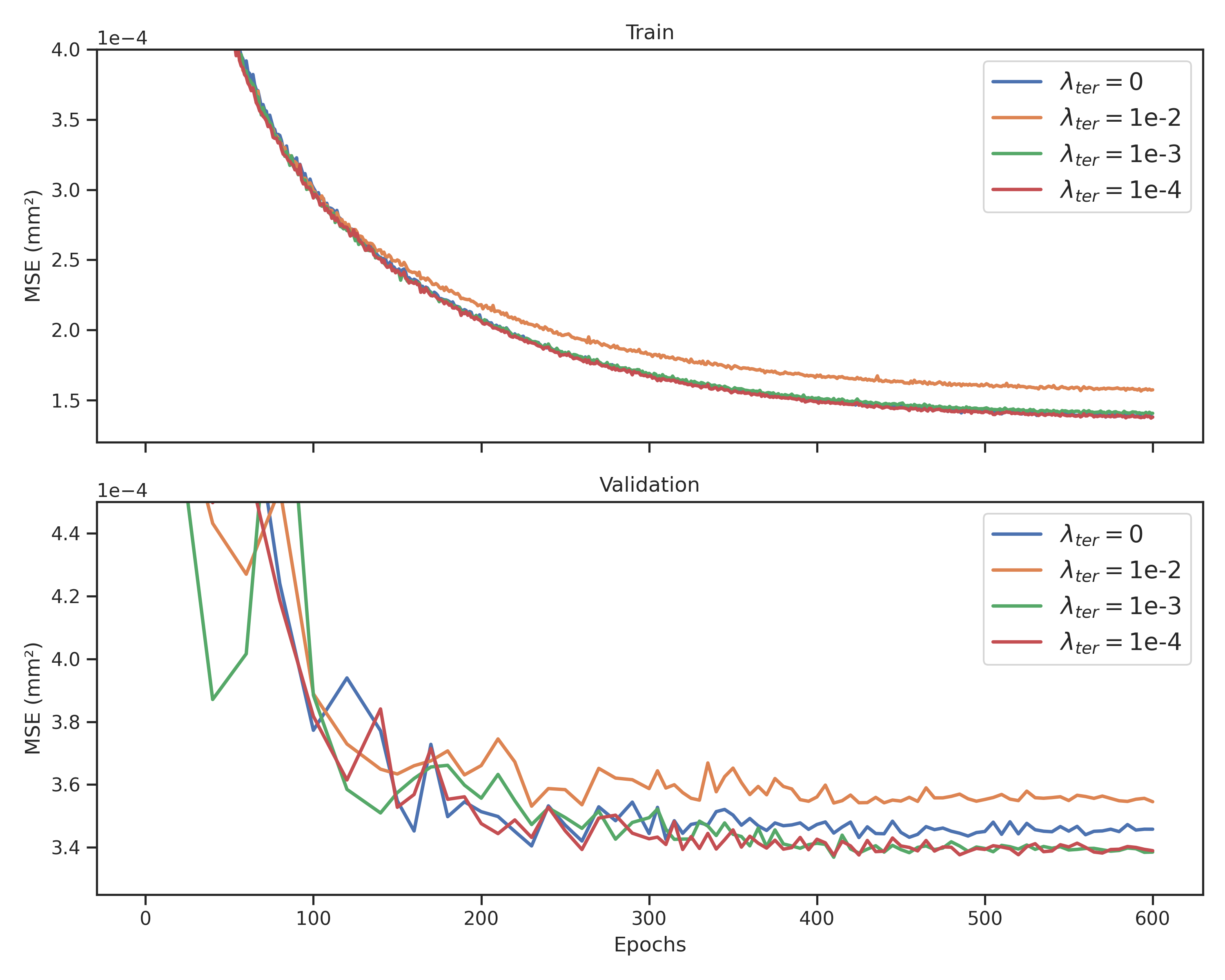}
    \caption{MSE values throughout training and validation for volumetric meshes, exploring different configurations of $\lambda_{ter}$, with values measured in the \textit{relative positional space}. Noticeably, $\lambda_{ter} = 1\text{E-}2$ (\textcolor{black}{Orange}) shows a high-performance decay for both train and validation curves. Smaller intervals were used for loss recording as training progressed.}
    \label{fig:losses_volumetric}
\end{figure}

A closer examination of the training dynamics, as illustrated in Figure \ref{fig:losses_volumetric}, reinforces the benefits of using small values of $\lambda_{ter}$. These values result in improved validation performance in terms of vertex MSE without substantial fluctuations in the training curves. On the contrary, the highest regularisation strength ($\lambda_{ter} = 1\text{E-}2$) leads to decreased performance in both training and validation. 

\begin{figure}[t!]
    \centering
    \includegraphics[width=\linewidth]{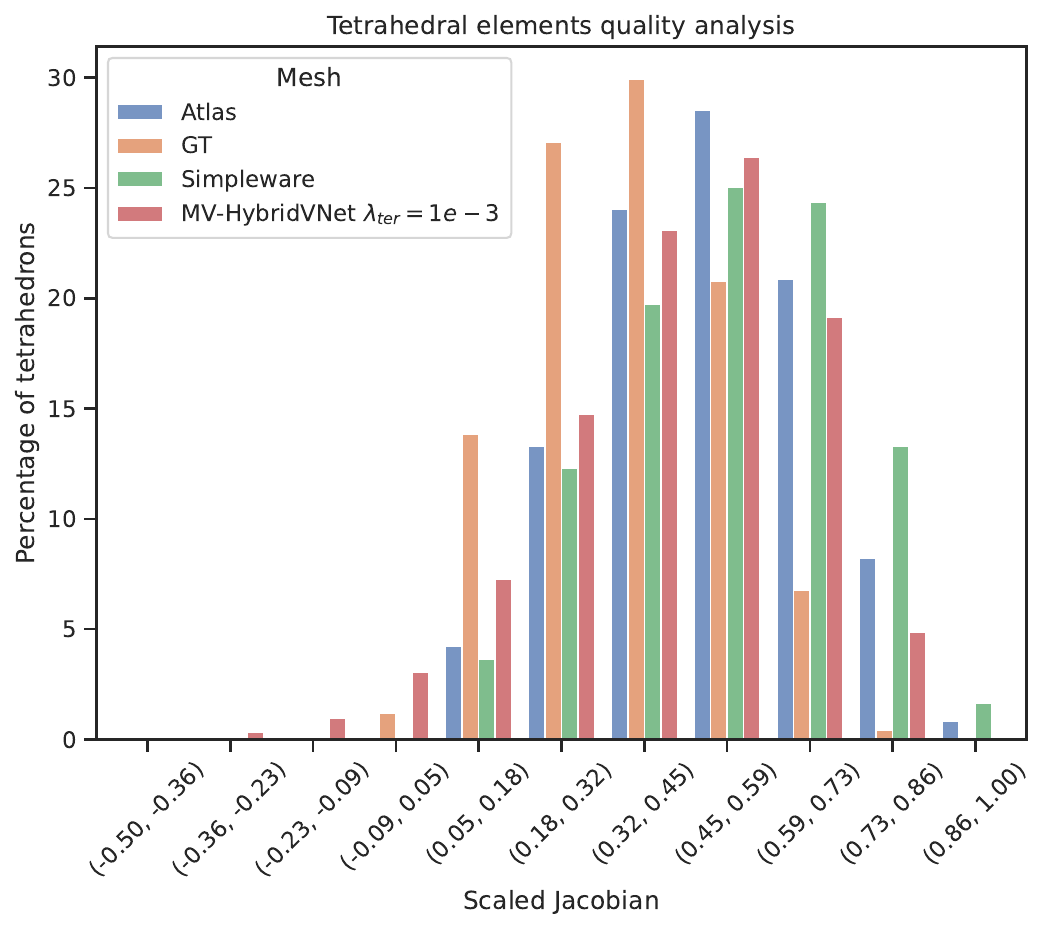}
    \caption{Histogram of tetrahedral mesh quality using scaled Jacobian values. The x-axis represents the scaled Jacobian values, and the y-axis shows the percentage of tetrahedral elements within each range.}%
    \label{fig:histogram}
\end{figure}

\textcolor{black}{For an in-depth analysis of mesh quality, Table \ref{table:volumetric_qualities} provides a comprehensive overview of scaled Jacobian values under different conditions}, comparing our approaches with volumetric atlases, ground-truth meshes, and a subset of surface meshes converted to volumetric meshes using Simpleware's ScanIP \cite{Simpleware}. \textcolor{black}{The analysis reveals that ground-truth meshes used for training, obtained through atlas registration, exhibit poor quality characteristics with a mean Jacobian of 0.355, and contain degenerated tetrahedra as evidenced by the \textcolor{black}{minimum Jacobian value} of -0.207. Our regularized approach demonstrates significant quality improvements, with $\lambda_{ter} = 1\text{E-}2$ achieving a mean Jacobian of 0.501, surpassing the average element quality of both ground truth and atlas meshes, and approaching the quality of Simpleware-generated meshes (0.524), despite being trained on imperfect data.} Our regularised models surpass ground-truth elements in terms of quality, beginning from the 25\% quartile and onwards, for $\lambda_{ter} = 1\text{E-}3$ and higher, underscoring that the regularisation loss significantly enhances mesh quality. Figure \ref{fig:histogram} visually summarises this improvement, positioning our method competitively with Simpleware meshes, except for a small number of elements, potentially due to the original low quality of the ground truth.

\input{table6}

Overall, our model demonstrates competitive results compared to the conventional approach of directly converting surface to volumetric meshes. Moreover, it addresses a challenge posed by direct conversion, where degenerate triangles can obstruct the creation of volumetric meshes, affecting approximately 10\% of cases in our experiments. When comparing the time required for generating a volumetric mesh, Simpleware's ScanIP procedure consumes approximately 6 minutes on average for each mesh, employing the same configuration as used in the atlas generation procedure. In contrast, our approach requires less time for generating the vertex set of volumetric meshes. When executed on an NVIDIA A100-SXM4 GPU, it accomplishes this task in just 0.04 seconds for each set of CMR images during the forward pass, resulting in a substantial speed improvement. Even in cases where GPU computing is unavailable, when running on an Intel(R) Core(TM) i7-7700 CPU operating at 3.60GHz, the forward pass requires only 5 seconds on average, providing a significant acceleration.

\subsection{Limitations}

\textcolor{black}{Although our method exhibits promising results when compared with state-of-the-art methods, it is not free from limitations. One of its main limitations is the requirement in training time of ground-truth meshes with fixed topology, i.e. all subject meshes should have the same number of vertices and faces. While we acknowledge this limitation, it is worth noting that this is well aligned with computational anatomy applications, where correspondences between subjects are essential for statistical analysis. This is particularly true in cardiac imaging, where most existing meshes are created using atlas-based or point-based approaches, which naturally generate fixed topology meshes and correspondences between subjects. Nevertheless, in future research we plan to extend our framework to handle varying mesh topologies, enabling its use in new application scenarios where such meshes are not currently available.}

\textcolor{black}{Another limitation is related to the lack of fine-grained evaluation of our model in highly anomalous or pathological cases. Here we considered the UK Biobank CMR Dataset \cite{petersen2015uk}, because this has become a standard benchmark for evaluating cardiac mesh extraction methods (see for example \cite{kong2023,chen2021shape}). Even though most of the patients correspond to healthy controls, this dataset also includes a smaller proportion of samples with myocardial infarction (about 150 patients) that were considered during the evaluation. Illustrative examples of the resulting meshes are included in Figure \ref{fig:qualitative}. Nonetheless, in future work we will conduct a comprehensive performance analysis of HybridVNet focusing on highly anomalous cases, to gain deeper insight into its robustness when handling outliers and complex pathological anatomies.}

\textcolor{black}{Additionally, it is important to note that even though the cardiac atlas mesh used in the study (available from \cite{rodero2021}) includes the base of the aorta and pulmonary artery, and all cardiac valve planes in addition to all four cardiac chambers, the quality of the meshes inferred for aorta and pulmonary artery is not as reliable as the rest of the structures. This is due to the fact that pulmonary and aortic arteries were not included in the original manual contours; instead, these structures were inferred from the rest of the delineated anatomy during registration.}

\textcolor{black}{Finally, since our meshes are based on a generative decoder, we cannot guarantee by construction that the different cardiac structures will not intersect. However, several factors make this highly unlikely. First, our method preserves the topological connectivity in the atlas of Rodero et al. \cite{rodero2021}, ensuring shared nodes at interfaces, which inherently mitigates intersection risks. Second, the use of regularisation losses, both for surface and volumetric meshes, prevents extreme deformations that could lead to self-intersections. Additionally, the ground-truth meshes used for training were generated via a mesh-to-contour registration process that incorporated both 3D coordinates and normal vectors, further ensuring accurate correspondence and preventing overlap \cite{XIA2022102498}. As a result, our trained model inherits the non-intersecting property of the ground-truth meshes, while our combined approach of anatomical connectivity and geometric regularisation offers strong safeguards against mesh irregularities.}

\section{Conclusions}

This study introduces HybridVNet, a novel method for directly generating surface and tetrahedral meshes from images. Our comprehensive experiments and evaluations reveal that HybridVNet significantly enhances mesh accuracy and versatility compared to state-of-the-art point distribution models that depend on linear PCA component decoding. In particular, integrating short- and long-axis views \textcolor{black}{improves the accuracy of the generated meshes}. HybridVNet stands out for its efficiency and speed, substantially reducing vertex set generation time compared to conventional approaches, a precious trait for large-scale processing such as in studies on the UK Biobank.

The generic nature of HybridVNet opens doors to broader applications in medical image analysis, with potential extensions to tasks such as cortical surface reconstruction from brain magnetic resonance images. \textcolor{black}{Future work will direct efforts toward improving the quality of the tetrahedral ground truth used for model training, as the current training dataset contains suboptimal elements due to the atlas registration process. This could potentially lead to even better mesh quality in the predicted results.}

\section*{Acknowledgments}

AFF acknowledges support from the Royal Academy of Engineering under the RAEng Chair in Emerging Technologies (INSILEX CiET1919/19) and the ERC Advanced Grant – UKRI Frontier Research Guarantee (INSILICO EP/Y030494/1). EF acknowledges support from Nvidia for the donation of GPU computing, the Argentinian Agencia Nacional de Promoción de la Investigación, el Desarrollo Tecnológico y la Innovación (PICT PRH 2019-0009), and Universidad Nacional del Litoral (CAID project).

\bibliographystyle{IEEEtran}
\bibliography{bibtex.bib}

\end{document}

%% file: table1.tex
\begin{table*}[ht]
\resizebox{\linewidth}{!}{
\begin{tabular}{p{0.22\linewidth}p{0.18\linewidth}p{0.2\linewidth}p{0.29\linewidth}p{0.29\linewidth}}
\toprule
\textbf{Mesh generation} &
  \textbf{Process Step} &
  \textbf{Algorithms involved} &
  \textbf{Hyper-parameters} &
  \textbf{Comments} \\ \midrule
\multirow{5}{*}{Common pipeline \cite{KIM2018161, VAANANEN201919, pak2024robustautomatedcalcificationmeshing, chen2021shape}} &
  Segmentation from an 3D image set with dense coverage &
  Convolutional neural networks, multi-atlas segmentation &
  CNN architecture selection, neural network training hyperparameters, number of atlases and atlas selection &
  Explicitly limited by voxel-size and by the rectangular shape of each voxel \\ \cline{2-5} 
 &
  Postprocessing of densely-covering segmentation masks &
  Artifact removal, smoothing, resampling, hole filling &
  Size of the artifacts to remove, kernel sizes for smoothing, interpolation method, manual hole filling &
  Semi automatic procedure, open problem of generating anatomically plausible segmentations \\ \cline{2-5} 
 &
  Generation of surface mesh from segmentation masks &
  Marching cubes, marching tetrahedra &
  Grid resolution, isovalue, interpolation method &
  Grid dependency, staircase effect, limited handling of noise, triangle quality \\ \cline{2-5} 
 &
  Postprocessing surface meshes &
  Laplacian smoothing, mesh decimation, hole filling, normal smoothing, topological cleaning, edge smoothing &
  Smoothing factor and iterations, target vertex count, maximum hole size, hole filling method, normal computation method, cleaning methods &
  Manual intervention required \\ \cline{2-5} 
 &
  Volumetric mesh generation from surface meshes &
  Delaunay tetrahedralization, quality control, mesh optimization &
  Tetrahedralization algorithm parameters, surface mesh constraints, boundary preservation thresholds, quality metric selection and threshold, smoothing parameters and optimization objectives &
  Slow and with manual intervention required \\ \midrule
\multirow{3}{*}{Surface mesh pipeline \cite{kong2023}} &
  Surface mesh extraction directly from 3D images &
  Convolutional neural networks mixed with point-distribution models &
  CNN architecture, point distribution model selection, neural network training hyperparameters &
  Can incorporate anatomical information in the PDM model and generate meshes with no topological artifacts \\ \cline{2-5} 
 &
  Mesh postprocessing &
  Laplacian smoothing, normal smoothing, edge smoothing &
  Smoothing factor and iterations, normal computation method &
  Manual intervention may be required \\ \cline{2-5} 
 &
  Generation of volumetric meshes from surface meshes &
  Delaunay tetrahedralization, quality control, mesh optimization &
  Tetrahedralization algorithm parameters, surface mesh constraints, boundary preservation thresholds, quality metric selection and threshold, smoothing parameters and optimization objectives &
  Slow and with manual intervention required \\ \midrule
Proposed volumetric mesh pipeline &
  Extraction of surface or volumetric meshes directly from raw images &
  \textbf{Hybrid graph convolutional neural networks (proposed)} &
  CNN encoder and decoder architectures, neural network training hyper-parameters &
  Incorporates topological information via the adjacency matrix, learns anatomical embeddings in the hybrid network bottleneck, and direct specification of mesh quality via regularization terms at training time. \\ \bottomrule
\end{tabular}}
\caption{Comparison of Mesh Generation Pipelines}
\label{tab:mesh_generation}
\end{table*}

%% file: table2.tex
\begin{table*}[ht!]
\centering
\begin{tabular}{cccccccc}
\hline
\multicolumn{1}{l}{\multirow{2}{*}{}} & \multirow{2}{*}{\textbf{Metrics}} & \textbf{MCSI-Net SAX} & \multicolumn{2}{c}{\textbf{HybridVNet}} & \textbf{MCSI-Net SAX-LAX} & \multicolumn{2}{c}{\textbf{MV-HybridVNet}} \\ \cline{3-8} 
\multicolumn{1}{l}{} &  & \textbf{Cropped} & \multicolumn{1}{c}{\textbf{Full Image}} & \textbf{Cropped} & \textbf{Cropped} & \multicolumn{1}{c}{\textbf{Full Image}} & \textbf{Cropped} \\ \hline
\multirow{3}{*}{\textbf{LV Endo}} & DC $\uparrow$ & 0.87 (0.05) & \multicolumn{1}{c}{0.89 (0.05)} & \textbf{0.90 (0.04)} & 0.88 (0.05) & \multicolumn{1}{c}{0.90 (0.04)} & \textbf{0.91 (0.04)} \\  
 & HD \textcolor{black}{(mm)} $\downarrow$ & 5.13 (1.97) & \multicolumn{1}{c}{4.48 (1.32)} & \textbf{4.08 (1.22)} & 4.74 (1.75) & \multicolumn{1}{c}{4.22 (1.22)} & \textbf{3.89 (1.18)} \\  
 & MCD \textcolor{black}{(mm)} $\downarrow$ & 1.93 (0.83) & \multicolumn{1}{c}{1.67 (0.55)} & \textbf{1.49 (0.49)} & 1.86 (0.79) & \multicolumn{1}{c}{1.55 (0.51)} & \textbf{1.39 (0.46)} \\ \hline
\multirow{3}{*}{\textbf{LV Myo}} & DC $\uparrow$ & 0.76 (0.09) & \multicolumn{1}{c}{0.80 (0.06)} & \textbf{0.83 (0.05)} & 0.78 (0.08) & \multicolumn{1}{c}{0.81 (0.05)} & \textbf{0.84 (0.04)} \\  
 & HD \textcolor{black}{(mm)} $\downarrow$ & 5.31 (1.98) & \multicolumn{1}{c}{4.71 (1.36)} & \textbf{4.23 (1.27)} & 4.75 (1.76) & \multicolumn{1}{c}{4.40 (1.26)} & \textbf{3.96 (1.23)} \\  
 & MCD \textcolor{black}{(mm)} $\downarrow$ & 1.97 (0.95) & \multicolumn{1}{c}{1.71 (0.56)} & \textbf{1.49 (0.51)} & 1.86 (0.82) & \multicolumn{1}{c}{1.57 (0.52)} & \textbf{1.35 (0.46)} \\ \hline
\multirow{3}{*}{\textbf{RV Endo}} & DC $\uparrow$ & 0.85 (0.06) & \multicolumn{1}{c}{0.85 (0.05)} & \textbf{0.86 (0.05)} & 0.85 (0.06) & \multicolumn{1}{c}{0.86 (0.05)} & \textbf{0.87 (0.05)} \\  
 & HD \textcolor{black}{(mm)} $\downarrow$ & 7.11 (2.78) & \multicolumn{1}{c}{6.97 (2.31)} & \textbf{6.44 (2.19)} & 7.06 (2.64) & \multicolumn{1}{c}{6.79 (2.23)} & \textbf{6.13 (2.23)} \\  
 & MCD \textcolor{black}{(mm)} $\downarrow$ & 2.34 (0.98) & \multicolumn{1}{c}{2.10 (0.64)} & \textbf{1.90 (0.57)} & 2.27 (0.95) & \multicolumn{1}{c}{1.99 (0.59)} & \textbf{1.76 (0.59)} \\ \hline
\end{tabular}
\caption{\textcolor{black}{Quantitative evaluation of surface mesh segmentation ($n=1,200$). Arrows indicate whether metrics improve with higher ($\uparrow$) or lower ($\downarrow$) values. Bold values indicate statistically significant improvements ($p<0.05$, independent t-test) compared to baseline models.}}
\label{surf_seg_metrics}
\end{table*}

%% file: table3.tex
\begin{table*}[t!]
\centering
\begin{tabular}{rrrrrr}
\hline
\multirow{2}{*}{\textbf{Subpart}} & \multirow{2}{*}{\textbf{Metric}} & \multicolumn{2}{c}{\textbf{Full SAX Image}} & \multicolumn{2}{c}{\textbf{Cropped SAX Image}} \\ 
\cline{3-4} \cline{5-6} 
 &  & \multicolumn{1}{c}{\textbf{HybridVNet}} & \textbf{MV-HybridVNet} & \multicolumn{1}{c}{\textbf{HybridVNet}} & \textbf{MV-HybridVNet} \\ \hline
\multirow{2}{*}{\textbf{Full Mesh}} & MAE \textcolor{black}{(mm)} $\downarrow$ & \multicolumn{1}{r}{2.56 (0.62)} & \textbf{2.26 (0.55)} & \multicolumn{1}{r}{2.43 (0.59)} & \textbf{2.18 (0.54)} \\ 
 & MSE \textcolor{black}{(mm$^{2}$)} $\downarrow$ & \multicolumn{1}{r}{12.20 (7.11)} & \textbf{9.29 (5.48)} & \multicolumn{1}{r}{11.27 (6.69)} & \textbf{8.80 (5.31)} \\ 
\hline
\multirow{2}{*}{\textbf{LV}} & MAE \textcolor{black}{(mm)} $\downarrow$ & \multicolumn{1}{r}{1.90 (0.57)} & \textbf{1.79 (0.55)} & \multicolumn{1}{r}{1.75 (0.54)} & \textbf{1.70 (0.54)} \\ 
 & MSE \textcolor{black}{(mm$^{2}$)} $\downarrow$ & \multicolumn{1}{r}{6.23 (4.28)} & \textbf{5.60 (4.03)} & \multicolumn{1}{r}{5.35 (3.83)} & 5.11 (3.67) \\ 
\hline
\multirow{2}{*}{\textbf{RV}} & MAE \textcolor{black}{(mm)} $\downarrow$ & \multicolumn{1}{r}{2.18 (0.64)} & \textbf{2.08 (0.60)} & \multicolumn{1}{r}{2.00 (0.58)} & 1.97 (0.59) \\ 
 & MSE \textcolor{black}{(mm$^{2}$)} $\downarrow$ & \multicolumn{1}{r}{8.39 (5.64)} & \textbf{7.69 (4.93)} & \multicolumn{1}{r}{7.12 (4.84)} & 7.04 (4.72) \\ 
\hline
\multirow{2}{*}{\textbf{LA}} & MAE \textcolor{black}{(mm)} $\downarrow$ & \multicolumn{1}{r}{2.90 (1.00)} & \textbf{2.37 (0.78)} & \multicolumn{1}{r}{2.84 (0.99)} & \textbf{2.30 (0.77)} \\ 
 & MSE \textcolor{black}{(mm$^{2}$)} $\downarrow$ & \multicolumn{1}{r}{15.40 (13.73)} & \textbf{10.07 (9.74)} & \multicolumn{1}{r}{14.88 (13.29)} & \textbf{9.58 (9.24)} \\ 
 \hline
\multirow{2}{*}{\textbf{RA}} & MAE \textcolor{black}{(mm)} $\downarrow$ & \multicolumn{1}{r}{3.07 (0.96)} & \textbf{2.57 (0.76)} & \multicolumn{1}{r}{2.98 (0.93)} & \textbf{2.51 (0.80)} \\ 
 & MSE \textcolor{black}{(mm$^{2}$)} $\downarrow$ & \multicolumn{1}{r}{17.46 (13.65)} & \textbf{12.00 (9.42)} & \multicolumn{1}{r}{16.67 (13.13)} & \textbf{11.75 (10.16)} \\ 
\hline
\multirow{2}{*}{\textbf{AORTA}} & MAE \textcolor{black}{(mm)} $\downarrow$ & \multicolumn{1}{r}{2.66 (0.93)} & \textbf{2.37 (0.84)} & \multicolumn{1}{r}{2.56 (0.89)} & \textbf{2.34 (0.83)} \\ 
 & MSE \textcolor{black}{(mm$^{2}$)} $\downarrow$ & \multicolumn{1}{r}{13.17 (11.05)} & \textbf{10.24 (8.71)} & \multicolumn{1}{r}{12.38 (10.52)} & \textbf{10.04 (8.43)} \\ 
 \hline
\end{tabular}
\caption{\textcolor{black}{Comparison of single-view (HybridVNet) versus multi-view (MV-HybridVNet) approaches on full and cropped SAX images ($n=1,200$). Arrows ($\uparrow$,$\downarrow$) indicate the desired direction for each metric. Bold values indicate statistically significant improvements ($p<0.05$, independent t-test) between models within each image type.}}
\label{surf_mesh_metrics}
\end{table*}

%% file: table4.tex
\begin{table*}[h!]
\centering
\begin{tabular}{lccc}
\hline
\textbf{Methods} & \textbf{CD (mm)} & \textbf{HD (mm)} & \textbf{Inference Time (ms)} \\ \hline
PointNet+                & 13.03 (2.96)  & 17.04 (3.57)  & $<$ 0.1 \\
PU-Net                   & 12.15 (2.88)  & 15.74 (3.37)  & $<$ 0.1 \\
Pixel2mesh               & 19.38 (5.54)  & 16.20 (3.30)  & $<$ 0.1 \\
CPD                      & 12.10 (6.63)  & 13.05 (7.04)  & 37.45 \\
GMMREG                   & 20.90 (7.18)  & 18.57 (3.04)  & 60.90 \\
MR-Net                   & 4.39 (1.48)   & 6.89 (1.88)   & $<$ 0.1 \\ \hline
HybridVNet (Surface)     & \textbf{4.13 (1.16)}   & \textbf{5.17 (1.02)}   & $<$ 0.1 \\ 
HybridVNet (Volumetric)  & \textbf{2.80 (2.40)}   & \textbf{6.09 (2.09)}   & $<$ 0.1 \\ \hline
\end{tabular}
\caption{\textcolor{black}{Comparison of segmentation-to-mesh methods on bi-ventricular mesh generation ($n=1,914$). Bold values indicate statistically significant improvements ($p<0.05$, independent t-test) over the best baseline (MR-Net). Baseline results were considered as reported in \cite{chen2021shape}.} }
\label{tab:comparison_seg_to_mesh}
\end{table*}


%% file: table5.tex
\begin{table*}[t!]
    \centering
\begin{tabular}{cccccc}
\hline
\multicolumn{1}{l}{\multirow{2}{*}{}} & \multicolumn{1}{l}{\multirow{2}{*}{Metrics}} & \multicolumn{4}{c}{\textbf{MV-HybridVNet}}                                                                                                                         \\ \cline{3-6} 
\multicolumn{1}{l}{}                  & \multicolumn{1}{l}{}                         & \multicolumn{1}{c}{$\lambda_{ter} = 0$} & \multicolumn{1}{c}{$\lambda_{ter} = 1\text{E-}4$} & \multicolumn{1}{c}{$\lambda_{ter} = 1\text{E-}3$} & $\lambda_{ter} = 1\text{E-}2$ \\ \hline
\multirow{2}{*}{\textbf{Mesh}}          & MAE $\downarrow$                              & \multicolumn{1}{c}{2.08 (0.63)}           & \multicolumn{1}{c}{2.07 (0.64)}              & \multicolumn{1}{c}{2.04 (0.61)*}     & 2.11 (0.61)*              \\  
                                        & MSE $\downarrow$                              & \multicolumn{1}{c}{8.25 (6.14)}           & \multicolumn{1}{c}{8.22 (6.12)}              & \multicolumn{1}{c}{7.93 (5.63)}    & 8.39 (6.00)     \\          \hline
\multirow{3}{*}{\textbf{LV Endo}}       & DC $\uparrow$                                 & \multicolumn{1}{c}{\textbf{0.90 (0.04)}}  & \multicolumn{1}{c}{\textbf{0.90 (0.04)}}     & \multicolumn{1}{c}{0.90 (0.05)}              & 0.88 (0.05)              \\ \ 
                                        & HD \textcolor{black}{(mm)} $\downarrow$                               & \multicolumn{1}{c}{\textbf{4.36 (1.22)}}  & \multicolumn{1}{c}{\textbf{4.32 (1.24)}}     & \multicolumn{1}{c}{\textbf{4.41 (1.35)}}              & 5.21 (1.42)              \\ \ 
                                        & MCD \textcolor{black}{(mm)} $\downarrow$                              & \multicolumn{1}{c}{\textbf{1.52 (0.46)}}  & \multicolumn{1}{c}{\textbf{1.51 (0.49)}}     & \multicolumn{1}{c}{1.58 (0.54)}              & 1.89 (0.62)              \\ \hline
\multirow{3}{*}{\textbf{LV Myo}}        & DC $\uparrow$                                 & \multicolumn{1}{c}{\textbf{0.78 (0.04)}}  & \multicolumn{1}{c}{\textbf{0.78 (0.04)}}     & \multicolumn{1}{c}{0.76 (0.05)}              & 0.74 (0.06)              \\ \ 
                                        & HD \textcolor{black}{(mm)} $\downarrow$                               & \multicolumn{1}{c}{5.27 (1.47)}           & \multicolumn{1}{c}{\textbf{4.98 (1.40)}}     & \multicolumn{1}{c}{5.17 (1.50)}              & 5.30 (1.57)              \\ \ 
                                        & MCD \textcolor{black}{(mm)} $\downarrow$                              & \multicolumn{1}{c}{1.86 (0.61)}           & \multicolumn{1}{c}{\textbf{1.81 (0.64)}}     & \multicolumn{1}{c}{1.95 (0.72)}              & 1.96 (0.77)              \\ \hline
\multirow{3}{*}{\textbf{RV Endo}}       & DC $\uparrow$                                 & \multicolumn{1}{c}{\textbf{0.85 (0.06)}}           & \multicolumn{1}{c}{\textbf{0.86 (0.05)}}     & \multicolumn{1}{c}{\textbf{0.85 (0.05)}}              & 0.85 (0.06)              \\ \ 
                                        & HD \textcolor{black}{(mm)} $\downarrow$                               & \multicolumn{1}{c}{7.22 (2.76)}           & \multicolumn{1}{c}{\textbf{6.97 (2.54)}}     & \multicolumn{1}{c}{7.38 (2.67)}              & 7.55 (2.80)              \\ \ 
                                        & MCD \textcolor{black}{(mm)} $\downarrow$                              & \multicolumn{1}{c}{\textbf{2.05 (0.64)}}           & \multicolumn{1}{c}{\textbf{2.02 (0.63)}}     & \multicolumn{1}{c}{2.09 (0.64)}              & 2.13 (0.69)              \\ \hline
\multirow{5}{*}{\textbf{Mesh Quality}} & Scaled Jacobian $\uparrow$ & \multicolumn{1}{c}{0.22 (0.23)} & \multicolumn{1}{c}{0.23 (0.23)} & \multicolumn{1}{c}{0.43 (0.21)} & \textbf{0.50 (0.31)} \\ \ 
 & Aspect Ratio $\downarrow$ & \multicolumn{1}{c}{52.46 (16302.31)} & \multicolumn{1}{c}{24.49 (2212.50)} & \multicolumn{1}{c}{8.16 (861.52)} & \textbf{4.57 (702.27)} \\ \
 & Mean Ratio $\uparrow$ & \multicolumn{1}{c}{0.38 (0.35)} & \multicolumn{1}{c}{0.38 (0.35)} & \multicolumn{1}{c}{0.66 (0.25)} & \textbf{0.69 (0.40)} \\ \
 & Skewness $\downarrow$ & \multicolumn{1}{c}{0.69 (0.15)} & \multicolumn{1}{c}{0.68 (0.15)} & \multicolumn{1}{c}{0.54 (0.16)} & \textbf{0.43 (0.15)} \\ \
 & Shape Quality $\uparrow$ & \multicolumn{1}{c}{0.43 (0.26)} & \multicolumn{1}{c}{0.44 (0.26)} & \multicolumn{1}{c}{0.67 (0.21)} & \textbf{0.74 (0.24)} \\ \hline
\end{tabular}
    \caption{\textcolor{black}{Evaluation of volumetric mesh segmentation and quality metrics ($n=1,200$). Arrows ($\uparrow$,$\downarrow$) next to each metric indicate whether higher or lower values are better. Bold values indicate the best performing methods exhibiting significant differences with respect to the non-bold values ($p<0.05$, independent t-test), but no significant differences among themselves. Asterisk (*) marks significant differences between columns.}}
    \label{table:volumetric}
\end{table*}


%% file: table6.tex
\begin{table*}[ht]
    \centering
    \begin{tabular}{ccccccccccc}
    \hline
    \multicolumn{2}{c}{}                                                       & \textbf{Mean} & \textbf{Std} & \textbf{Min} & \textbf{Max} & \textbf{1\%} & \textbf{5\%} & \textbf{25\%} & \textbf{50\%} & \textbf{75\%} \\ \hline
    \multicolumn{1}{c}{\multirow{3}{*}{\textbf{Reference Meshes}}} & Atlas                  & 0.491         & 0.174        & 0.092        & 0.984        & 0.115        & 0.194        & 0.367         & 0.494         & 0.617         \\  
    \multicolumn{1}{c}{}                                           & Ground Truth           & 0.355         & 0.156        & -0.207       & 0.838        & 0.04         & 0.103        & 0.238         & 0.353         & 0.47          \\  
    \multicolumn{1}{c}{}                                           & Simpleware             & 0.524         & 0.185        & 0.064        & 0.992        & 0.128        & 0.202        & 0.387         & 0.535         & 0.667         \\ \hline
    \multicolumn{1}{c}{\multirow{4}{*}{\textbf{MV-HybridVNet}}}  & $\lambda_{ter} = 0$    & 0.222         & 0.225        & -0.759       & 0.876        & -0.327       & -0.144       & 0.065         & 0.219         & 0.384         \\  
    \multicolumn{1}{c}{}                                           & $\lambda_{ter} = 1\text{E-}4$ & 0.229         & 0.23         & -0.771       & 0.871        & -0.337       & -0.151       & 0.068         & 0.231         & 0.397         \\  
    \multicolumn{1}{c}{}                                           & $\lambda_{ter} = 1\text{E-}3$ & 0.433         & 0.206        & -0.719       & 0.904        & -0.138       & 0.059        & 0.307         & 0.457         & 0.585         \\  
    \multicolumn{1}{c}{}                                           & $\lambda_{ter} = 1\text{E-}2$ & 0.501         & 0.309        & -0.931       & 0.943        & -0.681       & -0.298       & 0.434         & 0.577         & 0.688         \\ \hline
    \end{tabular}
    \caption{Quality assessment of volumetric mesh elements. Values represent scaled Jacobian, with higher values indicating better quality.}
    \label{table:volumetric_qualities}
\end{table*}